# Correspondence Truth and Quantum Mechanics

## Vassilios Karakostas[1]

**Abstract** The logic of a physical theory reflects the structure of the propositions referring to the behaviour of a physical system in the domain of the relevant theory. It is argued in relation to classical mechanics that the propositional structure of the theory allows truth-value assignment in conformity with the traditional conception of a correspondence theory of truth. Every proposition in classical mechanics is assigned a definite truth value, either 'true' or 'false', describing what is actually the case at a certain moment of time. Truth-value assignment in quantum mechanics, however, differs; it is known, by means of a variety of 'no go' theorems, that it is not possible to assign definite truth values to all propositions pertaining to a quantum system without generating a Kochen-Specker contradiction. In this respect, the Bub-Clifton 'uniqueness theorem' is utilized for arguing that truth-value definiteness is consistently restored with respect to a determinate sublattice of propositions defined by the state of the quantum system concerned and a particular observable to be measured. An account of truth of contextual correspondence is thereby provided that is appropriate to the quantum domain of discourse. The conceptual implications of the resulting account are traced down and analyzed at length. In this light, the traditional conception of correspondence truth may be viewed as a species or as a limit case of the more generic proposed scheme of contextual correspondence when the non-explicit specification of a context of discourse poses no further consequences.

**Keywords** Correspondence theory · Propositional truth · Quantum mechanics · Kochen-Specker theorem · Maximal determinate sublattice · Contextual interpretation · Truth-conditions

## 1 Truth as Correspondence: The Traditional Framework

In investigations concerning the problem of truth in the physical sciences, the correspondence theory of truth has frequently been thought of as the most eminent. Although the correspondence theory admits various different formulations, the core of any correspondence theory is the idea that a proposition is true if and only if it corresponds to or matches reality. The classical version of the theory describes this relationship as a correspondence to the facts about the world (e.g., Burgess et al. 2011, pp. 70-72). If so, then adopting a correspondence theory of truth amounts to endorsing instances of the following scheme:

*Correspondence to facts* [CF]: The proposition that *P* is true if and only if *P* corresponds to a fact.

Alternatively, if one construes the notion of a 'fact' in terms of the weaker notion of an obtaining 'state of affairs', as in an Austin-type theory, then, [CF] is re-expressed as follows:[2]

---

[1] Department of Philosophy and History of Science, Faculty of Sciences, University of Athens, University Campus, Athens 157 71, Greece; Email: karakost@phs.uoa.gr

> *Correspondence to states of affairs* [CS]: The proposition that *P* is true if and only if there is a state of affairs *X* such that *P* corresponds to *X* and *X* obtains.

The useful feature of states of affairs is that they refer to something that can be said to obtain or fail to obtain, to be the case or not to be the case, to be a fact or fail to be a fact, that is, they exist even when they are not concretely manifested or realized.

The logical factor involved in the traditional view of correspondence truth implies that wherever there is a true proposition there is a fact (or state of affairs) stated by it and wherever a fact (or state of affairs) a possible true proposition which states it. This seems uncontroversial, as being reflected into the biconditional connective 'if and only if' of the preceding alethic schemes, but taken in that way the theory is not illuminating. For, "corresponds to the facts" can function merely as a synonym, as an alternative extended way of saying "is true" (e.g., Lewis 2001). It is not clear, therefore, that anything substantial can be said about the correspondence relation, except that it is the relation in which a proposition stands to the world when the proposition is true; nor is it clear that the relevant facts, or states of affairs, can be specified except as those which make a particular proposition true.

If one wishes to generalize this and inquire into the necessary and sufficient conditions for a proposition to be true, it is difficult to go beyond the truism that if a proposition is true then the relevant fact is as the proposition says it is. Then, using a popular example in the philosophy literature, the proposition 'snow is white' is true because it corresponds to the fact that snow is white. Thus, the fact that makes a proposition true is a restatement of the proposition itself. Facts are merely re-expressions of the propositions they make true.

In this sense, in the traditional correspondence notion of truth, the truth-conditions, namely, the worldly conditions that make a proposition *P* true are simply given by *P* itself. It is natural to think, however, that if an account of truth in terms of correspondence with facts is not to be idle, one must deploy a notion of fact and of correspondence that would allow us to go further than the trivial equivalence between "it is true that *P*" and "it is a fact that *P*" (e.g., Engel 2002). A correspondence theory, fully worthy of the name, must go on to articulate an explanation of the correspondence relation that is more complex, and thus not amenable to the immediate restatement reply. In addition, there must provide a genuine account of facts as special kinds of entities that can be candidates for the relationship of truth-making.

---

[2] Classical accounts of a correspondence theory of truth, such as Russell's and Austin's accounts, are both included in a recent collection of essays concerning the nature of truth edited by Lynch (2001, pp. 17, 25). See, also, Candlish et al. (2007, pp. 241, 266).

In particular, when examining the functioning of correspondence truth within physical science, it should be clear that it requires an understanding not just of the logical form of correspondence per se, but of the specific field of knowledge in which the correspondence relation is realized. Within this domain of inquiry, the primary question asked should be about what objects there are in the world, and what properties and relations are instantiated by these objects. Then, truth in terms of correspondence may be appropriately understood as *property instantiation*, in the following sense: if $P$ is a true proposition, then $P$ attributes some property to some object of the relevant domain. In this respect, the criticism exerted to the usual conception of correspondence truth does not imply the impossibility of a genuine correspondence account of truth, but, rather, by taking into account the specificities of the physics discourse, it attempts to determine the applicability and, if need be, to adjust appropriately the content of a correspondence account of truth when applied to the propositional language of fundamental physics (see, especially, Section 4).

It is worthy to note in this connection that the traditional theory of truth as correspondence, regardless of its exact formulation, has frequently been associated with the view that truth is radically non-epistemic (e.g., Devitt 2001, p. 606). This means that:

> *Radically non-epistemic conception of truth*: The truth (or falsity) of a proposition is entirely independent of anyone's cognitive capacities, beliefs, theories, conceptual schemes, and so on.

According to this non-epistemic conception, the truthmakers of propositions (namely, facts or actual states of affairs) are totally independent of human conceptualization and thus are among the intrinsic furniture of a mind-independent reality. Consequently, in order to say whether a given proposition is true or false one will need to have access to the way things are independently of our ability to verify or confirm or justify or even test this proposition. That is, to determine the truth of a proposition it seems that we must either compare it to raw, un-conceptualized reality or somehow have a privileged, direct and unproblematic access to reality.[3] Henceforth, on this conception, facts or actual states of affairs, forming the object-end of the truthmaking relation, are considered as being already *given*, as being *completely autonomous* in themselves, or as residing in the world *purely extensionally*, that is, in a manner independent of our worldviews and particular discursive practices and contexts.

---

[3] McDermid (1998), Blackburn (1996, p. 85), and Bonjour (1985, p. 7) all offer versions of the so-called 'comparison problem' as the fundamental obstacle faced by a correspondence theory of truth.

In this sense, the view of a radically non-epistemic conception of truth incorporates the following transcendence condition:

> *Transcendence condition*: The truth of a proposition transcends our possible knowledge of it, or its evidential basis; it is empirically unconstrained.

As a consequence of the preceding condition, even if it is impossible to produce a framework on which we may ascertain the truth value of a proposition this does not imply that the proposition does not possess any such value. It always has one. It is either determinately true or determinately false independently of any empirical evidence or cognitive means by which we may establish which value it is. The possession of truth values is therefore entirely independent of our means of warranting their assignment.

The transcendence condition, no doubt, attempts to capture the realist intuition that a proposition cannot be claimed true or false in virtue of its knowability or justifiability. For, a proposition may be true without being justified, and vice versa. Agreed! But what does the transcendence thesis, in its totality, really presuppose, especially when viewed within the traditional framework of correspondence truth? It presupposes the existence of a 'platonic' universe of true propositions, entirely independent of our ability in having access to it, and, henceforth, elements of this ideal world, in this case, propositions, possess determinate truth values entirely independent of our capability in forming justified convictions about them. In other words, it is important to realize that this thesis does not simply aim to establish an objective basis or attribute a non-epistemic character to the notion of truth — that, for instance, the content of declarative propositions is rendered true (or false) on the basis of worldly conditions and not on some relevant beliefs of ours — but this particular conception tends to be so radically non-epistemic that at the end leads to a notion of truth with absolutely no epistemic features. Be that as it may, I shall argue immediately below that the propositional structure of classical mechanics allows truth-value assignments in conformity with the usual traditional conception of a correspondence account of truth.

## 2 Propositional Truth in Classical Mechanics

In classical mechanics a system $S$ with $n$ degrees of freedom is described by a phase space $\Omega_S$ with $2n$ coordinates $\{q_i, p_i\}$ which correspond to generalized position and momentum coordinates. The state of $S$ at any temporal moment $t$ is represented by a point $X_t = \{q_i(t), p_i(t)\}$ of $\Omega_S$. Physical quantities are represented by real-valued functions on the phase space, e.g., the position $q$ of a mass point is a function $q: \Omega_S \rightarrow R^3$. Physical properties — namely, values of various physical quantities of the system — are represented by Borel subspaces $\Omega_S^A$, $\Omega_S^B$, … of $\Omega_S$ and will be denoted by $P(A)$, $P(B)$, …, respectively. Hence, a property is represented by a characteristic function $P(A): \Omega_S \rightarrow \{0, 1\}$ with $P(A)(X) = 1$ if $X \in \Omega_S^A$ and $P(A)(X) = 0$ if $X \notin \Omega_S^A$. We say that the characteristic function takes the value 1 or the property $P(A)$ pertains to system $S$ at time $t$ if the state of $S$ is represented by a point lying in the corresponding subset ($X_t \in \Omega_S^A$), and that $P(A)$ does not pertain to $S$ if the state of the system is represented by a point outside this subset ($X_t \notin \Omega_S^A$). In terms of propositions $P_A$, $P_B$, … this means that a proposition $P_A$ is true if the property $P(A)$ pertains to $S$, and false otherwise. That is, the proposition $P_A$ asserting that 'system $S$ acquires the property $P(A)$', or equivalently, that 'the value $a$ of some physical quantity $A$ of $S$ lies in a certain range of values $\Delta$' ('$a \in \Delta$'), is true if and only if the associated property $P(A)$ obtains. In the propositional structure of classical mechanics, each point in phase space, representing a classical state of a given system $S$, defines a truth-value assignment to the subsets representing the propositions. Each subset to which the point belongs represents a true proposition or a property that is instantiated by the system. Likewise, each subset to which the point does not belong represents a false proposition or a property that is not instantiated by the system. Thus, every possible property of $S$ is selected as either occurring or not; equivalently, every corresponding proposition pertaining to $S$ is either true or false.

Hence, for present purposes, the really essential thing about the mode of representation of classical systems is that the algebra of properties or propositions of a classical mechanical system is isomorphic to the lattice of subsets of phase space, a Boolean lattice $L_B$ that can be interpreted semantically by a 2-valued truth-function. This means that to every proposition $P \in L_B$ one of the two possible truth values 1 (true) and 0 (false) can be assigned through the associated characteristic function; equivalently, any proposition is either true or false (*tertium non datur*) (e.g., Dalla Chiara et al. 2004, p. 21). Thus, the propositions of a classical system are *semantically decidable*, independently of any perceptual evidence or cognitive means by which we may verify or falsify them.

From a physical point of view this is immediately linked to the fact that classical physics views objects-systems as bearers of determinate properties. Specifically, classical physical systems are taken to obey a so-called 'possessed values' or 'definite values' principle that may be succinctly formulated as follows:[4]

> *Definite values principle*: Any classical system is characterized, at each instant of time, by definite values for *all* physical quantities pertaining to the system in question.

That is, classical properties (values of physical quantities) are considered as being *intrinsic* to the system, as being possessed by the system itself. They are independent of whether or not any measurement is attempted on them and their definite values are independent of one another as far as measurement is concerned. Successive measurements of physical quantities, like position and momentum that define the state of a classical system, can be performed to any degree of accuracy and the results combined can completely determine the state of the system before and after the measurement interaction, since its effect, if not eliminable, takes place continuously in the system's phase space and is therefore predictable in principle. Hence, during the act of measurement a classical system conserves its identity; measurement does not induce any qualitative changes on the state of the measured system. Thus, the principle of value-definiteness implicitly incorporates the following assumption of non-contextuality:

> *Non-contextuality*: If a classical system possesses a property (value of a physical quantity), then it does so independently of any measurement context, i.e., independently of *how* that value is eventually measured.

This means that the properties possessed by a classical system depend in no way on the relations obtaining between it and a possible experimental or measurement context used to bring these properties about. If a classical system possesses a given property, it does so independently of possessing other values pertaining to other experimental arrangements. All properties pertaining to a classical system are simultaneously determinate and potentially available to the system, regardless of our means of exploring and warranting their assignment. Accordingly, the propositions of a classical system are considered as possessing determinate truth values — they are either determinately true or determinately false — *prior* to and *independent* of any actual investigation of the states of affairs the propositions denote; that is, classical mechanical propositions possess investigation-independent truth values, thus capturing the radically non-epistemic character of a traditional correspondence account of truth. Truth-value definiteness is conceived in virtue of a stable and well-defined external reality which serves as the implicit referent of every proposition, so that it bears no further consequences in avoiding specifying the exact domain of reference. Consequently, in a classical universe of discourse, there is supposed to exist implicitly an Archimedean standpoint from which the totality of facts may be logically evaluated.

---

[4] The principle of value-definiteness has variously been called in the literature as, for instance, "the determined value assumption" in Auletta (2001, pp. 21, 105).

## 3  Propositional Truth in Quantum Mechanics

On the standard (Dirac-von Neumann) interpretation of quantum theory, the elementary propositions pertaining to a quantum mechanical system form a non-Boolean lattice, $L_H$, isomorphic to the lattice of closed linear subspaces or corresponding projection operators of a Hilbert space. Thus, a proposition pertaining to a quantum system is represented by a projection operator $P$ on the system's Hilbert space $H$ or, equivalently, it is represented by the linear subspace $H_P$ of $H$ upon which the projection operator $P$ projects (e.g., Rédei 1998, Sect. 4.2). Since each projection operator $P$ on $H$ acquires two eigenvalues 1 and 0, where the value 1 can be read as 'true' and 0 as 'false', the proposition 'a system $S$ in state $|\psi\rangle$ has the property $P(A)$' is said to be true if and only if the corresponding projection operator $P_A$ obtains the value 1, that is, if and only if $P_A |\psi\rangle = |\psi\rangle$. Accordingly, the state $|\psi\rangle$ of the system lies in the associated subspace $H_A$ which is the range of the operator $P_A$, i.e., $|\psi\rangle \in H_A$. In such a circumstance, the property $P(A)$ pertains to the quantum system $S$. Otherwise, if $P_A |\psi\rangle = 0$ and, hence, $|\psi\rangle \in \perp H_A$ (subspace completely orthogonal to $H_A$), the counter property $\neg P(A)$ pertains to $S$, and the proposition is said to be false. It might appear, therefore, that propositions of this kind have a well-defined truth value in a sense analogous to the truth-value assignment in classical mechanics.

There is, however, a significant difference between the two situations. Unlike the case in classical mechanics, for a given quantum system, the propositions represented by projection operators or Hilbert space subspaces are not partitioned into two mutually exclusive and collectively exhaustive sets representing either true or false propositions.

As already pointed out, only propositions represented by subspaces that contain the system's state are assigned the value 'true' (propositions assigned probability 1 by $|\psi\rangle$), and only propositions represented by spaces orthogonal to the state are assigned the value 'false' (propositions assigned probability 0 by $|\psi\rangle$) (Dirac 1958, pp. 46-47; von Neumann 1955, pp. 213-217). Hence, propositions represented by subspaces that are at some non-zero or non-orthogonal angle to the unit vector $|\psi\rangle$ or, more appropriately, to the ray representing the quantum state are not assigned any truth value in $|\psi\rangle$. These propositions are neither true nor false; they are assigned by $|\psi\rangle$ a probability value different from 1 and 0; thus, they are *undecidable* or *indeterminate* for the system in state $|\psi\rangle$ and the corresponding properties are taken as indefinite. Suppose, for instance, we are referring to the spin property of a simple system, say an electron, whose spin in a certain direction may assume only two possible values: either $+\frac{1}{2}\hbar$ ('up') or $-\frac{1}{2}\hbar$ ('down'). Now, the spin in the *z*-direction, $S_z$, and the spin in the *y*-direction, $S_y$, represent two incompatible observables that cannot be measured simultaneously. Suppose an electron in state $|\psi\rangle$ satisfies the proposition '$S_z$ is up', so that $|\psi\rangle$ is an eigenvector of $S_z$. Consequently, both propositions '$S_y$ is up' and '$S_y$ is down', represented by subspaces at some angle other than 0 or $\pi/2$ to the state $|\psi\rangle$, shall be indeterminate. In such a circumstance, it cannot be asserted meaningfully that $S_y$ possesses a specific value; any of the possible values of $S_y$ can neither be regarded as manifested by the system in state $|\psi\rangle$, nor as excluded for the system. In particular, the proposition '$S_y$ is up' (or '$S_y$ is down') cannot be assigned a meaningful truth value.

This kind of semantic ambiguity far from signifying a perplexing feature peculiar to the spin property of one-half particles constitutes an inevitable consequence of the Hilbert-space structure of quantum mechanics demonstrated rigorously, for the first time, by Kochen-Specker's (1967) theorem. According to this, for any quantum system associated to a Hilbert space of dimension higher than two, there does not exist a 2-valued, truth-functional assignment $h: L_H \rightarrow \{0, 1\}$ on the set of closed linear subspaces, $L_H$, interpretable as quantum mechanical propositions, preserving the lattice operations and the orthocomplement. In other words, the gist of the theorem, when interpreted semantically, asserts the impossibility of assigning definite truth values to *all* propositions pertaining to a physical system at any one time, for any of its quantum states, without generating a contradiction. What are, therefore, the maximal sets of subspaces of $L_H$ or the maximal subsets of propositions that can be taken as simultaneously determinate, that is, as being assigned determinate (but perhaps unknown) truth values in an overall consistent manner?

3.1  Maximal Sets of Simultaneously Determinate Quantum Mechanical Propositions

In this respect, we employ the Bub-Clifton so-called 'uniqueness theorem' (Bub 2009; Bub & Clifton 1996). Consider, to this end, a quantum system $S$ represented by an $n$-dimensional Hilbert space whose state is represented by a ray or one-dimensional projection operator $D = |\psi\rangle\langle\psi|$ spanned by the unit vector $|\psi\rangle$ on $H$. Let $A$ be an observable of $S$ with $m \leq n$ distinct eigenspaces $A_i$, while the rays $D_{A_i} = (D \vee A_i^\perp) \wedge A_i$, $i = 1, \ldots, k \leq m$, denote the non-zero projections of the state $D$ onto these eigenspaces. Then, according to the Bub-Clifton theorem, the unique maximal sublattice of the lattice of projection operators or subspaces, $L_H$, representing the propositions that can be determinately true or false of the system $S$, is given by

$$L_H(\{D_{A_i}\}) = \{P \in L_H : D_{A_i} \leq P \text{ or } D_{A_i} \leq P^\perp, \forall i, i = 1, \ldots, k\}.$$

The sublattice $L_H(\{D_{A_i}\}) \subset L_H$ is generated by (i) the rays $D_{A_i}$, the non-zero projections of $D$ onto the $k$ eigenspaces of $A$, and (ii) all the rays in the subspace $(D_{A_1} \vee D_{A_2} \vee \ldots \vee D_{A_k})^\perp = (\vee D_{A_i})^\perp$ orthogonal to the subspace spanned by the $D_{A_i}$, for $i = 1, \ldots, k$. Since the $D_{A_i}$ are orthogonal, they are compatible and generate a Boolean sublattice of $L_H$. So

$$(\vee D_{A_k})^\perp = (D_{A_1})^\perp \wedge (D_{A_2})^\perp \wedge \ldots (D_{A_k})^\perp.$$

Effectively, the system's Hilbert space is partitioned into $k$-orthogonal subspaces corresponding to a partition of the spectrum of $A$ into $k$ distinct eigenspaces. Hence,

$$L_H(\{D_{A_k}\}) = L_H(D_{A_1}) \cap L_H(D_{A_2}) \cap \ldots L_H(D_{A_k}),$$

since each $L_H(D_{A_i})$, $i = 1, \ldots, k$, is generated by the ray $D_{A_i}$ and all the rays in the subspaces $(D_{A_i})^\perp$ orthogonal to $D_{A_i}$. The set of maximal (non-degenerate) observables associated with $L_H(\{D_{A_k}\})$ includes any maximal observable with $k$ eigenvectors in the directions $D_{A_i}$, $i = 1, \ldots, k$. The set of non-maximal observables includes any non-maximal observable that is a function of one of these maximal observables. Thus, all the observables whose eigenspaces are spanned by rays in $L_H(\{D_{A_k}\})$ are determinate, given the system's state $D$ and $A$.

Identifying such maximal determinate sets of observables amounts, in effect, to a consistent assignment of truth values to the associated propositions in $L_H(\{D_{A_k}\})$ of $L_H$, not to all propositions in $L_H$. $L_H(\{D_{A_k}\})$ represents the maximal subsets of propositions pertaining to a quantum system that can be taken as having simultaneously determinate truth values, where a truth-value assignment is defined by a 2-valued (or Boolean) homomorphism, $h: L_H(\{D_{A_k}\}) \to \{0,1\}$. If the system's Hilbert space $H$ is more than 2-dimensional, there are exactly $k$ 2-valued homomorphisms on $L_H(\{D_{A_k}\})$, where the $i^{th}$ homomorphism assigns to proposition $D_{A_i}$ the value 1 (i.e., true) and the remaining propositions in $L_H(\{D_{A_i}\})$, $i = 1, \ldots, k$, the value 0 (i.e., false). The determinate sublattice $L_H(\{D_{A_k}\})$ is maximal, in the sense that, if we add anything to it, lattice closure generates the lattice $L_H$ of all subspaces of $H$, and there are no 2-valued homomorphisms on $L_H$ (Bub 2009).

In fact, the Bub-Clifton determinate sublattice $L_H(\{D_{A_i}\})$ constitutes a generalization of the usual Dirac-von Neumann codification of quantum mechanics. On this standard position, an observable has a determinate value if and only if the state $D$ of the system is an eigenstate of the observable. Equivalently, the propositions that are determinately true or false of a system are the propositions represented by subspaces that either include the ray denoting the state $D$ of the system, or are orthogonal to $D$. Thus, the Dirac-von Neumann determinate sublattice can be formulated as

$$L_H(D) = \{P \in L_H: D \leq P \text{ or } D \leq P^\perp\}.$$

It is simply generated by the state $D$ and all the rays in the subspace orthogonal to $D$. If the system's Hilbert space $H$ is more than 2-dimensional, there is one and only 2-valued homomorphism on $L_H(D)$: the homomorphism induced by mapping the state $D$ onto 1 and every other ray orthogonal to $D$ onto 0. Apparently, the sublattice $L_H(D)$ for a particular choice of an observable $A$ in state $D$ forms a subset of Bub-Clifton's proposal $L_H(\{D_{A_i}\})$. The latter will only agree with $L_H(D)$ if $D$ is an eigenstate of $A$, for then the set $\{D_{A_i}\}$ consists of only $D$ itself. In general, the sublattice $L_H(\{D_{A_i}\})$ contains all the propositions in $L_H(D)$ that it makes sense to talk about consistently with $A$-propositions, namely propositions that are strictly correlated to the spectral projections of some suitable preferred observable $A$. From this perspective, the Dirac-von Neumann sublattice is obtained by taking $A$ as the unit (or identity) observable $I$. As Bub & Clifton (1996) rightly observe, however, there is nothing in the mathematical structure of Hilbert space quantum mechanics that necessitates the selection of the preferred determinate observable *A* as the unit observable *I*, whilst, in addition, this choice leads to von Neumann's account of quantum measurement resulting in a sequential regress of observing observers.

Then, the following question arises. What specifies the choice of a particular preferred observable *A* as determinate if $A \neq I$? The Bub-Clifton proposal allows, in effect, different choices for *A* corresponding to various different 'no collapse' interpretations of quantum mechanics, as for instance Bohm's (1952) hidden variable theory, if the privileged observable *A* is fixed as position in configuration space, or modal interpretations that exploit the bi-orthogonal decomposition theorem (e.g., Dieks et al. 1998). In them the preferred determinate observable is not always fixed but varies with the quantum state.

## 4   Contextual Semantics in Quantum Mechanics

4.1   Context-Dependent Assignment of Truth Values

In our view, if one wishes to stay within the framework of Hilbert space quantum mechanics and refrains from introducing additional structural elements, the most natural and immediate choice of a suitable preferred observable, especially, for confronting the problem of truth-value assignments, results in the determinateness of the observable to be measured. This is physically motivated by the fact that in the quantum domain one cannot assign, in a consistent manner, definite sharp values to *all* quantum mechanical observables pertaining to a microphysical object, in particular to pairs of incompatible observables, independently of the measurement context actually specified. In terms of the structural component of quantum theory, this is due to functional relationship constraints that govern the algebra of quantum mechanical observables, as revealed by the Kochen-Specker theorem alluded to above and its recent investigations (e.g., Cabello 2006; Kirchmair et al. 2009; Yu et al. 2012). In view of them, it is not possible, not even in principle, to assign to a quantum system definite non-contextual properties corresponding to *all* possible measurements. This means that it is not possible to assign a definite unique truth value to *every* single yes-no proposition, represented by a projection operator, independent of which subset of mutually commuting projection operators one may consider it to be a member. Hence, by means of a generalized example, if *A*, *B* and *E* denote observables of the same quantum system, so that the corresponding projection operator $A$ commutes with operators $B$ and $E$ ($[A, B] = 0 = [A, E]$), not however the operators $B$ and $E$ with each other ($[B, E] \neq 0$), then the result of a measurement of $A$ *depends* on whether the system had previously been subjected to a measurement of the observable $B$ or a measurement of the observable $E$ or in none of them. Thus, the value of the observable $A$ depends upon the set of mutually commuting observables one may consider it with, that is, the value of $A$ depends upon the selected set of measurements. In other words, the value of the observable $A$ cannot be thought of as pre-fixed, as being independent of the experimental context actually chosen, as specified, in our example, by the $\{A, B\}$ or $\{A, E\}$ frame of mutually compatible observables. Accordingly, the truth value assigned to the associated proposition '$a \in \varDelta$' — i.e., 'the value $a$ of the observable $A$ of system $S$ lies in a certain range of values $\varDelta$' — should be contextual as it depends on whether $A$ is thought of in the context of simultaneously ascribing a truth value to propositions about $B$, or to propositions about $E$. In fact, any attempt of simultaneously ascribing context-independent, sharp values to *all* observables of a quantum object forces the quantum statistical distribution of value assignment into the pattern of a classical distribution, thus leading directly to contradictions of the Greenberger-Horne-Zeilinger type (for a recent discussion see Greenberger 2009).

This state of affairs reflects most clearly the unreliability of the so-called 'definite values' principle of classical physics of Section 2, according to which, values of physical quantities are regarded as being possessed by an object independently of any measurement context. The classical underpinning of such an assumption is conclusively shown to be incompatible with the structure of the algebra of quantum mechanical observables. Whereas in classical physics, nothing prevented one from considering *as if* the phenomena reflected intrinsic properties, in quantum physics, even the *as if* is restricted. Indeed, quantum phenomena are not stable enough across series of measurements of non-commuting observables in order to be treated as direct reflections of invariable properties; the microphysical world seems to be sensitive to our experimental intervention.

Now, the selection of a particular observable to be measured necessitates also the selection of an appropriate experimental or measurement context with respect to which the measuring conditions remain intact. Formally, a measurement context $C_A(D)$ can be defined by a pair $(D, A)$, where, as previously, $D = |\psi\rangle\langle\psi|$ is an idempotent projection operator denoting the general initial state of system $S$ and $A = \sum_i a_i P_i$ is a self-adjoint operator denoting the measured observable. Of course, $C_A(D)$ is naturally extended to all commuting, compatible observables which, at least in principle, are co-measurable alongside of $A$. Then, in accordance with the Bub-Clifton theorem, given the state $D$ of $S$, $D$ restricted to the set of all propositions concerning $A$ is *necessarily* expressed as a weighted mixture $D_A = \sum_{i=1}^{k} |c_i|^2 |a_i\rangle\langle a_i|$ of determinate truth-value assignments, where each $|a_i\rangle$ is an eigenvector of $A$ and $|c_i| = |\langle \psi, a_i \rangle|$, $i = 1,..., k$. Since $D_A$ is defined with respect to the selected context $C_A(D)$, $D_A$ may be called a representative contextual state.[5] In other words, $D_A$ is a mixed state over a set of basis states that are eigenstates of the measured observable $A$, and it reproduces the probability distribution that $D$ assigns to the values of $A$. Thus, with respect to the representative contextual state $D_A$ the following conditions are satisfied:

i) Each $|a_i\rangle$ is an eigenvector of $A$. Thus, each quantum mechanical proposition $D_{A_i} \equiv P_{|a_i\rangle} = |a_i\rangle\langle a_i|$, $i = 1,..., k$, assigns in relation to $C_A(D)$ some well-defined value to $A$ (i.e., the eigenvalue $\alpha_i$ satisfying $A|a_i\rangle = \alpha_i |a_i\rangle$).

ii) Any eigenvectors $|a_i\rangle$, $|a_j\rangle$, $i \neq j$, of $A$ are orthogonal. Thus, the various possible propositions $\{P_{|a_i\rangle}\}$, $i = 1,..., k$, are mutually exclusive within $C_A(D)$. In this sense, the different orthogonal eigenstates $\{|a_i\rangle\}$, $i = 1,..., k$, correspond to different values of the measured observable $A$ or to different settings of the apparatus situated in the context $C_A(D)$.

iii) Each $|a_i\rangle$ is non-orthogonal to $D = |\psi\rangle\langle\psi|$. Thus, each proposition $P_{|a_i\rangle}$ whose truth value is not predicted with certainty is possible with respect to $C_A(D)$.

---

[5] In justifying from a physical point of view the aforementioned term, it is worthy to note that the state $D_A$, which results as a listing of well-defined properties or equivalently determinate truth-value assignments selected by a 2-valued homomorphism on $L_H (\{D_{A_i}\})$, may naturally be viewed as constituting a *state preparation* of system $S$ in the context of the preferred observable $A$ to be measured. Thus, the state $D_A$ should not be regarded as the final post-measurement state, reached after an $A$-measurement has been carried out on the system concerned. On the contrary, the contextual state represents here an alternative description of the initial state $D$ of $S$ by taking specifically into account the selection of a particular observable, and hence of a suitable experimental context, on which the state of the system under measurement can be conditioned. In other words, it provides a redescription of the measured system which is necessitated by taking specifically into account the context of the selected observable. For, it is important to realize that this kind of redescription is intimately related to the fact that both states $D$ and $D_A$ represent the *same* object system $S$, albeit in different ways. Whereas $D$ refers to a general initial state of $S$ independently of the specification of any particular observable, and hence, regardless of the determination of any measurement context, the state $D_A$ constitutes a conditionalization state preparation of $S$ with respect to the observable to be measured, while dropping all 'unrelated' reference to observables that are incompatible with such a preparation procedure.

It is evident, therefore, that the contextual state $D_A$ represents the set of all probabilities of events corresponding to quantum mechanical propositions $P_{|a_i\rangle}$ that are associated with the measurement context $C_A(D)$. In it the propositions $P_{|a_i\rangle}$ correspond in a one-to-one manner with disjoint subsets of the spectrum of the observable $A$ and hence generate a Boolean lattice of propositions.[6] Thus, the $P_{|a_i\rangle}$–propositions are assigned determinate truth values, in the standard Kolmogorov sense, by the state $D_A$.

It is instructive to note at this point that creating a preparatory Boolean environment $C_A(D)$ for a system $S$ in state $D$ to interact with a measuring arrangement does not determine which event will take place, but it does determine the *kind* of event that will take place. It forces the outcome, whatever it is, to belong to a certain definite Boolean sublattice of events for which the standard measurement conditions are invariant. Such a set of standard conditions for a definite kind of measurement constitutes a set of necessary and sufficient constraints for the occurrence of an event of the selected kind. This equivalently means, in relation to quantum theory, that the selection of an observable to be measured, by means of a corresponding preparation procedure, instantiates locally a physical context, which serves as a logical Boolean reference frame[7] for the individuation of events. It is probably one of the deepest insights of modern quantum theory that whereas the totality of all experimental/empirical events can only be represented in a globally non-Boolean structure, the acquisition of every single event depends on a locally Boolean context.

## 4.2 Contextual Account of Truth

In view of the preceding considerations, therefore, and in relation to philosophical matters, we propose a contextual account of truth that is compatible with the propositional structure of quantum theory by conforming to the following instance of the correspondence scheme:

---

[6] In fact, the determinate observable $A$ picks up a Boolean sublattice in $L_H(\{D_{A_i}\})$ which, in view of the Bub-Clifton theorem, is straightforwardly extended to $L_H(\{D_{A_i}\})$ itself. The latter comprises as determinate all observables whose eigenspaces are spanned by the rays $D_{A_i}$, given the system's state $D$ and $A$. These technicalities, however, bear no further significance for present purposes.

[7] Such a conceptual viewpoint has been also suggested in Davis (1977) and Takeuti (1978) and, recently, within a category-theoretic perspective of quantum theory, by Zafiris and Karakostas (2013).

*Contextual correspondence* [CC]: The proposition that *P*-in-*C* is true if and only if there is a state of affairs *X* such that (1) *P* expresses *X* in *C* and (2) *X* obtains, where *C* denotes, in general, the context of discourse, and specifically, in relation to the aforementioned quantum mechanical considerations, the experimental context $C_A(D)$ linked to the proposition $P \in L_H(\{D_{A_i}\})$ under investigation.

If, however, truth-value assignments to quantum mechanical propositions are context-dependent in some way as the scheme [CC] implies, it would appear, according to traditional thinking, that one is committed to antirealism about truth. In our opinion, this assumption is mistaken. The contextual account of truth suggested here is compatible with a realist conception of truth. Such an account essentially denies that there can be a 'God's-eye view' or an absolute Archimedean standpoint from which to state the totality of facts of nature. For, in relation to the microphysical world, there isn't a context-independent way of interacting with it. Any microphysical fact or event that 'happens' is raised at the empirical level only in conjunction with the specification of an experimental context that conforms to a set of observables co-measurable by that context (e.g., Svozil 2009).[8] In this respect, empirical access to the non-Boolean quantum world can only be gained by adopting a particular perspective, which is defined by a determinate sublattice $L_H(\{D_{A_i}\})$, or, in a more concrete sense, by the specification of an experimental context $C_A(D)$ that, in effect, selects a particular observable *A* as determinate. Within the context $C_A(D)$, the A-properties we attribute to the object under investigation have determinate values, but the values of incompatible observables, associated with incompatible (mutually exclusive) experimental arrangements, are indeterminate. Hence, at any temporal moment, there is no universal context that allows either an independent variation of the properties of a quantum object or a unique description of the object in terms of determinate properties. And this yields furthermore an explicit algebraic interpretation of the Bohrian notion of complementarity (a non-Copenhagen, of course), in so far as quantum mechanical properties obtain effectively determinate values — alternately, the associated propositions acquire determinate truth-value assignments — within a particular quasi-Boolean sub-structure $L_H(\{D_{A_i}\})$, whereas the underlying source of quantum mechanical 'strangeness' is located in the fact that they cannot be simultaneously realized or embedded within a single Boolean logical structure.

---

[8] It should be pointed out that Bohr already on the basis of his complementarity principle introduced the concept of a 'quantum phenomenon' to refer "exclusively to observations obtained under specified circumstances, including an account of the whole experiment" (Bohr 1963, p. 73). This feature of context-dependence is also present in Bohm's ontological interpretation of quantum theory by clearly putting forward that "quantum properties cannot be said to belong to the observed system alone and, more generally, that such properties have no meaning apart from the total context which is relevant in any particular situation. In this sense, this includes the overall experimental arrangement so that we can say that measurement is context dependent" (Bohm and Hiley 1993, p. 108).

Furthermore, the proposed account of truth, as encapsulated by the scheme [CC] of contextual correspondence, ought to be disassociated from an epistemic notion of truth. The reference to an experimental context in quantum mechanical considerations should not be viewed primarily as offering the evidential or verificationist basis for the truth of a proposition; it does not aim to equate truth to verification. Nor should it be associated with practices of instrumentalism, operationalism and the like; it does not aim to reduce theoretical terms to products of operational procedures. It rather provides the appropriate *conditions* under which it is possible for a proposition to receive consistently a truth value. Whereas in classical mechanics the conditions under which elementary propositions are claimed to be true or false are determinate independently of the context in which they are expressed, in contradistinction, the truth-conditions of quantum mechanical propositions are determinate within a context. On account of the Kochen-Specker theorem, there simply does not exist, within a quantum mechanical discourse, a consistent binary assignment of determinately true or determinately false propositions independent of the appeal to a context; propositional content seems to be linked to a context. This connection between referential context and propositional content means that a descriptive elementary proposition in the domain of quantum mechanics is, in a sense, incomplete unless it is accompanied by the specified conditions of an experimental context under which the proposition becomes effectively truth-valued (see, in addition, Karakostas 2012). In other words, the specification of the context is *part and parcel* of the truth-conditions that should obtain for a proposition in order the latter to be invested with a determinate (albeit unknown) truth value. Otherwise, the proposition is, in general, semantically undecidable. In the quantum description, therefore, the introduction of the experimental context is to select at any time $t$ a specific sublattice $L_H(\{D_{A_i}\})$ in the total non-Boolean lattice $L_H$ of propositions of a quantum system as co-definite; that is, each proposition in $L_H(\{D_{A_i}\})$ is assigned at time $t$ a definite truth value, 'true' or 'false', or equivalently, each corresponding property of the system either obtains or does not obtain. In effect, the specification of the context provides the necessary conditions whereby bivalent assignment of truth values to quantum mechanical propositions is *in principle* applicable. The obtainment of the conditions implies that it is possible for us to make, in an overall consistent manner, meaningful statements that the properties attributed to quantum objects are part of physical reality. This marks the fundamental difference between conditions for well-defined attribution of truth values to propositions and mere verification conditions.

This element also signifies the transition from the transcendence condition of the conventional correspondence theory of truth of Section 1 to a reflective-like transcendental reasoning of the proposed account of truth. That is, it signifies the transition from the uncritical qualification of truth values to propositions beyond the limits of experience and acknowledging them as being true or false *simpliciter*, to the demarcation of the limits of possible experience or to the establishment of pre-conditions which make possible the attribution of truth values to propositions. In the quantum description, therefore, the specification of the experimental context forms a *pre-condition* of quantum physical experience, which is necessary if quantum mechanics is to grasp empirical reality at all. In this respect, the specification of the context constitutes a methodological act preceding any empirical truth in the quantum domain and making it possible.

In closing this work, I wish to re-emphasize the fact that the proposed account of truth of contextual correspondence [CC] while preserves the realist intuition of the notion of correspondence truth, nonetheless, characterizes the makers of propositional truths as being context-dependent, if the world, in its microphysical dimension, is to be correctly describable. Truthmakers of quantum mechanical propositions, namely facts or actual states of affairs, are not pre-determined, pre-fixed; they are not 'out there' wholly unrestrictedly. And if facts are context-dependent, then whatever truths may be expressed about them must also be contextual. Truth contextuality follows naturally from the contextuality of facts. The truthmaking relationship is now established, not in terms of a raw un-conceptualized reality, as envisaged by the traditional scheme, but between a well-defined portion of reality as carved out by the experimental context and the propositional content that refers to the selected context. Such interdependence of propositional content and referential context is not by virtue of some meta-scientific principle or philosophical predilection, but by virtue of the microphysical nature of physical reality displaying a context-dependence of facts. For, as already argued, a quantum mechanical proposition is not true or false *simpliciter*, but acquires a determinate truth value with respect to a well-defined context of discourse as specified by the state of the quantum system concerned and the particular magnitude to be measured.

Thus, the conditions under which a proposition is true are *jointly* determined by the context in which the proposition is expressed and the actual microphysical state of affairs as projected into the specified context. In our approach, therefore, the reason that a proposition is true is because it designates an objectively existing state of affairs, albeit of a contextual nature. On the other hand, the traditional conception of correspondence truth, as exemplified either by the alethic scheme [CF] or [CS], alluded to in the introduction, and involving a direct context-independent relation between singular terms of propositions and definite autonomous facts of an external reality, may be viewed as a *species* or as a *limit case* of the more generic alethic scheme of contextual correspondence [CC], when the latter is applied in straightforward unproblematic circumstances where the non-explicit specification of a context of discourse poses no further consequences.

**Acknowledgements** For discussion and comments on previous versions, I thank participants at audiences in the Seventh European Conference of Analytic Philosophy (Milan) and Fourteenth International Congress of Logic, Methodology and Philosophy of Science (Nancy). I also acknowledge support from the research programme 'Thalis' co-financed by the European Union (ESF) and the Hellenic Research Council (project 70-3-11604).

## References

Auletta G (2001) Foundations and interpretation of quantum mechanics. World Scientific, Singapore

Blackburn S (1996) The Oxford dictionary of philosophy. Oxford University Press, Oxford

Bohm D (1952) A suggested interpretation of quantum theory in terms of "hidden variables", parts I and II. Phys Rev 85:166-179, 180-193

Bohm D, Hiley B (1993) The undivided universe: an ontological interpretation of quantum theory. Routledge, London

Bohr N (1963) Essays 1958-1962 on atomic physics and human knowledge. Wiley, New York

Bonjour L (1985) The structure of empirical knowledge. Harvard University Press, Cambridge, MA

Bub J (2009) Bub - Clifton theorem. In: Greenberger D, Hentschel K, Weinert F (eds) Compendium of quantum physics. Springer, Berlin, pp 84-86

Bub J, Clifton R (1996) A uniqueness theorem for "no collapse" interpretations of quantum mechanics. Stud Hist Phil Mod Phys 27:181-219

Burgess A, Burgess JP (2011) Truth. Princeton University Press, Princeton

Cabello A (2006) How many questions do you need to prove that unasked questions have no answers? Int J Quantum Inform 4:55-61

Candlish S, Damnjanovic N (2007) A brief history of truth. In: Jacquette D (ed) Philosophy of logic. Elsevier, Amsterdam, pp 227-323

Dalla Chiara M, Roberto G, Greechie R (2004) Reasoning in quantum theory. Kluwer, Dordrecht

Davis M (1977) A relativity principle in quantum mechanics. Int J Theor Phys 16:867-874

Devitt M (2001) The metaphysics of truth. In: Lynch M (ed) The nature of truth: classic and contemporary perspectives. MIT Press, Cambridge, MA, pp 579-611

Dieks D, Vermaas P (eds) (1998) The modal interpretation of quantum mechanics. Kluwer, Dordrecht

Dirac PAM (1958) Quantum mechanics, 4th edn. Clarendon Press, Oxford

Engel P (2002) Truth. Acumen, Chesham

Greenberger D (2009) GHZ (Greenberger-Horne-Zeilinger) theorem and GHZ states. In: Greenberger D, Hentschel K, Weinert F (eds) Compendium of quantum physics. Springer, Berlin, pp 258-263

Karakostas V (2012) Realism and objectivism in quantum mechanics. J Gen Phil Science 43:45-65

Kirchmair G, Zähringer F, Gerritsma R, Kleinmann M, Gühne O, Cabello A, Blatt R, Roos C (2009) State-independent experimental test of quantum contextuality. Nature 460:494-497

Kochen S, Specker E (1967) The problem of hidden variables in quantum mechanics. J Math Mech 17:59-87

Lewis D (2001) Forget about the 'correspondence theory of truth'. Analysis 61:275-280

Lynch M (ed) (2001) The nature of truth: classic and contemporary perspectives. MIT Press, Cambridge, MA

McDermid D (1998) Pragmatism and truth: the comparison objection to correspondence. Rev Metaphys 51:775-811

Rédei M (1998) Quantum logic in algebraic approach. Kluwer, Dordrecht

Svozil K (2009) Contexts in quantum, classical and partition logic. In: Engesser K, Gabbay D, Lehmann D (eds) Handbook of quantum logic and quantum structures: quantum logic. Elsevier, Amsterdam, pp 551-586

Takeuti G (1978) Two applications of logic to mathematics, part I: Boolean valued analysis. Publ Math Soc Japan 13, Iwanami and Princeton University Press, Tokyo and Princeton

Von Neumann J (1995) Mathematical foundations of quantum mechanics. Princeton University Press, Princeton

Yu S, Oh CH (2012) State-independent proof of Kochen-Specker theorem with 13 rays. Phys Rev Lett 108:030402

Zafiris E, Karakostas V (2013) A categorial semantic representation of quantum event structures. Foundations of Physics 43: 1090-1123